\documentstyle[preprint,prd,aps,12pt]{revtex}
\preprint{
$
\begin{array}{r}
\text{EP-CPTh.A000.0294} \\ \text{LAVAL-PHY-12-94}
\end{array}
$
}
\tighten

\begin{document}
\author{Luc Marleau\cite{Laval}}
\address{Centre de Physique Th\'eorique,\ \'Ecole\ Polytechnique\\
91128\ Palaiseau\ CEDEX,\ France}
\title{Uncertainties in testing QCD with \\ the photon structure function }
\date{February 1994 }
\draft
\maketitle
\begin{abstract}
We review the different approaches used to treat the photon structure
function. We suggest that despite some uncertainties it should remain
sensitive to $\Lambda .$
\end{abstract}

\input tcilatex

\section{Introduction}

As the years go by, QCD seems to remain the only serious candidate for a
theory of strong interactions. It has not encountered yet any unsurmountable
disagreement. But this is in part because, from a practical point of view,
it has a weakness, the non-perturbative sector
of the theory which is still largely out of theoretical reach.
Our inability to reach a non-perturbative
solution is indeed frustrating. Most, if not all, QCD tests has to deal with
some important non-perturbative component. This leaves us with two choices:
improve the non-perturbative methods (the final test of QCD may eventually
come from there), or isolate the non-perturbative effects, describing them
with phenomenologically inspired models or simple parameterizations, to base
the
discussion mainly on the calculable parts. Of course, the credibility of QCD
tests has suffered from this uncertainty especially with regard to the
dynamical properties, in hadron collisions for example.

There is actually one process which does not quite fit this description in
the sense that QCD predicts its shape, normalization as well as $Q^2$%
-evolution . This process is $\gamma \gamma ^{*}$
deep-inelastic scattering where a real photon structure function can be
extracted. The non-perturbative effects are not entirely absent though but
they contribute to higher orders in $\alpha _s$, the strong coupling
constant. One of the peculiarity of this process is that the structure
function has a strong dependence on $\alpha _s(Q)$ which makes it a good
candidate to the determination of the QCD scale parameter, $\Lambda $, as
was pointed out by Witten \cite{Witten}.

Two major contributions to the photon structure function can be identified:
the point-like part, which dominates in the high energy limit and provides
the predictions for shape and normalization, and the hadronic piece, on
which our knowledge is more limited. One of the problem in the past had to
do with the separation of these two pieces, which share cancelling
singularities. There are two prevailing attitude with that respect: One can
try to exploit as much as possible the QCD prediction by introducing a
proper regularization technique. This method introduce some uncertainties
which has be discussed in several papers \cite{Antoniadis}. On the other
hand, the
singularities can also be ignored if one is only interested in the $Q^2$%
-evolution of the structure function by means of a subtraction at a given
momentum scale $Q_0^2$ or $p_T\gg \Lambda ^2$. In this more
conservative approach however, the sensitivity to $\Lambda $ and the chance
to test QCD are almost completely lost.

In recent years, experimentalists have been led to analyze their data only
with respect to the second alternative. Clearly, deep-inelastic $ep$
scattering is better placed to provide a test of QCD's prediction for the $%
Q^2$-evolution of structure functions. In this talk, I will try to argue
that one should also look at the first approach and discuss briefly about
some misconceptions regarding both techniques.

\section{Witten's argument on sensitivity to $\Lambda $}

The QCD prediction for $F_{2,n}^\gamma (Q^2)$ (moments of the real photon
structure function) is
$$
F_{2,n}^\gamma (Q^2)\stackunder{Q^2\gg \Lambda ^2}{\simeq }\stackunder{\text{%
comes from the point-like part}}{\ \underbrace{\frac{a_n}{\alpha _s}+b_n}+(%
\text{higher orders in }\alpha _s)}
$$
where $\alpha _s\equiv \alpha _s(Q/\Lambda )$. The first two terms comes
from the so-called ``point-like'' part.

As mentionned above
the photon structure function can be separated into two pieces. First and
most important, is the point-like contribution. This part is essentially
due to the point-like nature of the photon and is absent in the structure
functions encountered in deep-inelastic $e$-$p$ scattering. The second
contribution is the so-called hadronic contribution and, hopefully, includes
all non-perturbative QCD effects that may contribute. What characterizes a
point-like object is that no matter how hard you hit it with a probe the
object shows no structure, in other words, the structure is independent of
any momentum scale. However, turning on strong interactions implies the
presence of at least two mass scales, the QCD $\Lambda $ parameter
and the $Q^2$ of the probe through the running coupling constant $\alpha
_s(Q)$. These are also the only scales possible in a ``point-like QCD''
process. Of course, strong interactions may be characterized by other
intermediate mass scale. Such a point of view is advocated in \cite{Field}
where the authors argue that some high-$p_T$ processes give indications of
an intermediate scale, $Q_0$, above which perturbative effects seems to
dominate. They then proceed by saying that only contributions with $Q^2\ge
Q_0^2$ participate to the point-like structure function. Needless to say
that this redefinition of ``point-like''
reduces this component with respect to the hadronic
component enough to jeopardize almost all hopes of testing QCD in $\gamma
\gamma $-scattering \cite{Bardeen}.

\section{The hadronic component and the singularities}

Let us rewrite the photon structure function in more details:%
$$
F_{2,n}^\gamma (Q^2)=\stackunder{\text{point-like}}{\underbrace{\frac{a_n}{%
\alpha _s}+b_n+\cdots }}+\stackunder{\text{hadronic}}{\underbrace{A_n\left[
\alpha _s\right] ^{d_n}}}
$$
Here, for simplicity we have omitted higher orders in $\alpha _s$ in the
point-like part (denoted here by $\left[ \cdots \right] $), higher orders
corrections to the hadronic part which would result in a power series
multiplying the last term and finally that the hadronic part do in fact get
three distinct contributions $A_n^i\left[ \alpha _s\right] ^{d_n^i}$ usually
denoted by $i=+,-,NS$ (non-singlet).

A number of comments are in order regarding this last expression to clarify
the separation between the point-like and hadronic parts: (1) The point-like
and hadronic part, which we denote from hereon by $F_{2,n}^{PL}(Q^2)$ and $%
F_{2,n}^{HAD}(Q^2)$ respectively, are both plagued by the
presence of singularities. These
are characterized by poles at given $n$'s for moments of the structure
function or equivalently by singular behavior in the low $x$ region in the $x
$-dependent structure function. Fortunately, each singularity in $%
F_{2,n}^{PL}(Q^2)$ is cancelled by a similar singularity in $%
F_{2,n}^{HAD}(Q^2).$ The total $F_{2,n}^\gamma (Q^2)$ is regular and
physically measurable as it should. (2) The point-like part should be in
principle independent of any scale since point-like objects look the same no
matter how hard one probes them. This is nearly the case here since the only
dependence on scale comes from $\alpha _s\,$ which it is caused by vacuum
polarization not by a structured object. Note also that $\alpha _s$ depends
on the only two scales defined in this problem $Q$ and $\Lambda $. In
principle, one can introduce arbitrary intermediate scales in this problem
but $F_{2,n}^\gamma (Q^2)$ is clearly independent on these scales. For
parameterization purposes, such scales are often introduced however one
should keep in mind that QCD gives no clear indications on which scales
should be preferred. (3) $A_n$ is the only non-perturbative (and in practice
non-calculable) object in this equation. Unlike in the virtual photon case,
it has no scale dependence. Its a pure function of $n$ (or $x$ when the
moments are inverted). (4) Finally, the $\left[ \alpha _s\right] ^{d_n}$
piece is the same as for the DIS on hadronic targets. Indeed, the $d_n$'s
depend on the same anomalous dimensions.

\section{Difficulty with introducing VDM into the picture}

Since the theoretical uncertainty regarding the photon structure
function comes from the hadronic part, it seems justified to use our
knowledge of the hadronic structure functions, more precisely through Vector
Meson Dominance (VDM) to get some insight on this question. There are indeed
several models which invoke VDM to justify their parameterization. This
should be done with great care since there are inherent difficulties in
identifying the hadronic part with VDM.

(i) It is in fact tempting to associate the hadronic piece of the structure
function directly with a VDM contribution%
$$
F_{2,n}^{HAD}(Q^2)\equiv A_n\left[ \alpha _s\right] ^{d_n}\stackrel{?}{=}%
F_{2,n}^{VDM}(Q^2)\equiv A_n^{VDM}\left[ \alpha _s\right] ^{d_n}
$$
but formally, this is not possible. $F_{2,n}^{HAD}(Q^2)$ must be singular to
cancel singularities in $F_{2,n}^{PL}(Q^2)$. The moments $a_n$, $b_n$,...are
singular at specific values of $n$. For example, near $n=2$, the moment
$b_n$ is ill-behaved:
$$
b_n\stackunder{n\rightarrow 2}{\sim }\frac b{n-2}
$$
which leads to a negative singular behavior in the small $x$ region (Regge
region)
$$
b(x)\stackunder{x\rightarrow 0}{\sim }-\frac bx.
$$
Clearly, $A_n$ must contain a similar singularities so that the total
$F_{2,n}^\gamma $ is free of singularities \cite{Bardeen}. For the $n=2$
case,%
$$
A_n\stackunder{n\rightarrow 2}{\sim }-\frac b{n-2}
$$
so $A_n\neq A_n^{VDM}$ since $A_n^{VDM}$ is regular for all $n$'s.

(ii) Instead one could think of separating the singular and regular pieces
of $A_n$ and identify the VDM\ contribution as follows:
$$
A_n=A_n|_{\text{singular}}+A_n^{VDM}.
$$

But there are some arbitrariness in the  definition of the singular term.
Adding any regular piece to the singular term gives an equally valid
definition for the singular term. On the other hand $A_n^{VDM}$ is
physically well defined and this implies that the singular term should be
uniquely defined. In fact, there seems to be no convincing way to define the
$A_n^{VDM}$ part in $A_n$. This kind of splitting can indeed lead to double
counting i.e. part of $A_n|_{\text{singular}}$ may be contributed by VDM.

(iii) There exist a popular alternative \cite{Field} which suggest to
consider the $Q^2$-evolution of $F_{2,n}^\gamma (Q^2)$ and assume that the
{\bf total }$F_{2,n}^\gamma (Q^2)$ is given at some lower $Q_0^2$ by $%
F_{2,n}^{VDM}(Q^2)$ alone. Of course once $F_{2,n}^\gamma (Q^2)$ is known at
some $Q_0^2$ it only remains to perform the $Q^2$-evolution and check if
this is in concordance with experiments. The choice of the scale $Q_0^2$ may
be physically motivated to some extent. For example one popular choice is
the scale at which QCD becomes perturbative (actually this scale is not so
well defined).

But this procedure has some degree of arbitrariness which comes in part from
the parameter $Q_0^2$. QCD gives no indications on the value of any
intermediate scale $Q_0^2.$ The only scale dependence of the predictions is
through the running coupling $\alpha _s.$ Furthermore $A_n$ which contains
the hadronic contribution is scale independent. It is also not clear that
the only contribution to the structure function below $Q_0^2$ origins from
VDM.

Since the only QCD prediction tested in this approach is the $Q^2$-evolution
(not the normalization) of $F_{2,n}^\gamma (Q^2)$, the sensitivity to the $%
\Lambda $ parameter is almost completely lost. Unfortunately, this approach
does ignore precious information coming from QCD on normalization and $%
\gamma \gamma $ processes are not the best place to test the $Q^2$%
-evolution. On the other hand, the above assumptions can be checked
experimentally. But one must be aware that an analysis of data according to
these models seems to have little chance of testing QCD itself. It will more
likely test the assumptions and the choice of $Q_0^2$.

(iv) There exists an even more cautious approach: Test only the $Q^2$%
-evolution by finding $F_{2,n}^\gamma (Q_0^2)$ experimentally at a given $%
Q_0^2.$ It has the advantage of being free of any ambiguity and model
independent. But it suffers from similar inconvenience namely loss of
sensitivity to $\Lambda ,$ loss of information from QCD and difficulty of
testing QCD from $\gamma \gamma $ processes due to small range of $Q^2$
spanned and lack of statistic.

\section{QCD regularization of $F_{2,n}^\gamma (Q^2)$}

There exists an approach which tries to exploit QCD theoretical predictions
as much as possible. The procedure consist in regularizing the singular
pieces from both the point-like and the hadronic part. Although it contains
some uncertainties, the procedure suggest how they may be controlled. Let us
rewrite%
$$
F_{2,n}^\gamma (Q^2)=F_{2,n}^{PL}(Q^2)+F_{2,n}^{HAD}(Q^2)
$$
where
$$
F_{2,n}^{PL}(Q^2)=\sum_{l=0}^\infty a_{l,n}\left[ \alpha _s\right] ^{l-1}\
\text{with\ }a_{l,n}=a_{l,n}^{\text{regular}}+\frac{a_l}{n-n_l}
$$
and
$$
F_{2,n}^{HAD}(Q^2)=A_n\left[ \alpha _s\right] ^{d_n}
$$

The regularization is similar to the one used for the virtual photon case
where the ambiguities are absent. It consists of replacing the cancelling
singular pieces by the following expression%
$$
\frac{a_l}{n-n_l}\left[ 1-\left( \lambda _l\alpha _s\right)
^{d_n+1-l}\right] \left[ \alpha _s\right] ^{l-1}=\frac{a_l\left[ \alpha
_s\right] ^{l-1}}{n-n_l}-\frac{a_l\lambda _l{}^{d_n+1-l}\left[ \alpha
_s\right] ^{d_n}}{n-n_l}
$$
where the first and second in the brakets on RHS comes from the point-like
and hadronic components respectively. The whole expression
is regular since $d_n+1-l\rightarrow 0$ as $n\rightarrow n_l$. (Note
that for the virtual photon case the cancellation, the parameter $\lambda _l$
is replaced by $\alpha _s^{-1}(p)$ where $p^2$ is the virtual photon mass).
The parameters $\lambda _l$ are in principle calculable from QCD and so are
well-defined objects. In practice however these are non-perturbative
parameters which escape as of now theoretical prediction.

What is interesting in the last expression is that it suggests
to use the $\lambda _l$ to parameterize the
total $A_n$ \cite{Antoniadis}:

$$
A_n=-\sum_{l=0}^\infty \frac{a_l}{n-n_l}\lambda _l{}^{d_n+1-l}
$$
In practice one only needs a finite number of $\lambda _l$'s to reach a
good fit
since the remaining regular pieces are suppressed by powers of $\alpha _s$
at large $l$'s. The second interesting fact is that the $\lambda _l$
parameters affects the region of small $x$ so these parameter can be fitted
from low $x$ data.

This QCD regularization is a physical (not an arbitrary mathematical \cite
{Storrow}) parameterization. The parameters can be determined from a
well-defined region of phase space and it corresponds to a convenient
parameterization.
Furthermore, it has been found that for a large range of values of $\lambda
_l,$ the $F_2^\gamma (x,Q^2)$ is still dominated by the regularized
point-like part at not too small $x.$ Therefore the sensitivity to $\Lambda $
is maintained in these regions. The procedure gives a well-defined approach
to control the theoretical uncertainties associated with the cancellations
of point-like and hadronic singularities.

\section{Conclusion}

In summary, a message to experimentalists: Two attitudes can prevail in
treating this problem. One may be cautious and, in view of uncertainties,
consider only the $Q^2$-evolution of the photon structure function. The
uncertainties are then parameterized either by fitting $F_2^\gamma (x,Q^2)$
to experimental data
 at  given $Q_0^2$ or by using some other assumption. In this
case, one cannot seriously expect to provide a conclusive test of QCD (it
more appropriate to look at DIS for $Q^2$-evolution) and
sensitivity to $\Lambda $
is almost completely lost. The second approach is
more opportunist for it makes use of as much information as QCD can provide.
In this case the uncertainties regarding the normalization can be
parameterized by a finite number of parameters. But most important, it seems
to indicate that a good determination of $\Lambda $ is still possible. In
any case, both parameterization should be analyzed until one is
experimentally excluded.

The author would acknowledge very useful discussions with G. Grunberg and
thank the Centre de Physique Th\'eorique (\'Ecole Polytechnique) where part
of this work was done, for their kind hospitality. This research was
supported by the Natural Science and Engineering Research Council of Canada
and by the Fonds pour la Formation de Chercheurs et l'Aide \`a la Recherche
du Qu\'ebec.



\begin{references}
\bibitem[\dag]{Laval}  On sabbatical leave from: D\'epartement de Physique,
Universit\'e Laval, Qu\'ebec, Canada, G1K 7P4.

\bibitem{Witten}  E. Witten, Nucl. Phys. B120, 189 (1977)

\bibitem{Bardeen}  W.A. Bardeen, iin Proceeding of the VI International
Workshop on Photon-Photon Collisions, Lake Tahoe, 1984, edited by R.L.
Lander (World Scientific, Singapore, 1985)

\bibitem{Field}  J.H. Field, F. Kapusta and L. Poggioli, Phys. Lett. B181,
362 (1986); Z. Phys. C36, 121 (1987). M. Gluck, E.Reya and A. Vogt, Phys.
Rev. D45, 3986 (1992); Phys. Rev. D46, 1973 (1992).

\bibitem{Drees}  M. Drees and K. Grassie, Z. Phys. C28, 451 (1985); H.
Abramowicz, K. Charcula and A. Levy, Phys. Lett. B269, 458 (1991).

\bibitem{Storrow}  J.H. Da Luz Vieira and J.K. Storrow, Z. Phys. C51, 241
(1991).

\bibitem{Antoniadis}  I. Antoniadis and G. Grunberg, Nucl. Phys. B213, 445
(1982); I. Antoniadis and L. Marleau, Phys. Lett. B161, 163 (1985).
\end{references}
\end{document}